\documentclass[twocolumn,english]{revtex4-1}
\usepackage[T1]{fontenc}
\usepackage{geometry}
\usepackage{amsmath}
\usepackage{graphicx}
\usepackage{babel}
\begin{document}
\title{Reentrant phase behavior in systems with density-induced tunneling}
\begin{abstract}
Open many body quantum systems play a paramount role in various branches
of physics, such as quantum information, nonlinear optics or condensed
matter. The dissipative character of open systems has gained a lot
of interest especially within the fields of quantum optics, due to
unprecedented stabilization of quantum coherence, and quantum information,
with its desire to control environmental degrees of freedom. We look
beyond the typical mechanism of dissipation associated with an external
source and show that strongly interacting many particle systems can
create quantum decoherence within themselves. We study a quantum bosonic
two-dimensional many body system with extended interactions between
particles. Analytical calculations show that the system can be driven
out of its coherent state, which is prevalent among commonly used
setups. However, we also observe a revival of the superfluid phase
within the same framework for sufficiently large interaction strength.
The breakdown of quantum coherence is inevitable, but can be misinterpreted
if one assumes improper coupling between the constituents of the many
particle system. We show an adequate path to retrieve physically relevant
results and consider its limitations. The system displays a natural
cutoff that enforces the breakdown of superfluidity.
\end{abstract}
\author{A. Krzywicka T. P. Polak}
\affiliation{Institute of Spintronics and Quantum Information, Faculty of Physics,
Adam Mickiewicz Univeristy in Pozna\'{n}}

\maketitle
The decoherence of quantum states due to coupling with external degrees
of freedom has gained a lot of interest \citep{ritsch2013,rai2015,klinder2015,zupancic2019,ferri2021,li2021,plenio1999,plenio2002}.
Especially enticing is the possibility of controlling how such systems
dissipate energy. For instance, quantum information processing relies
on precise control of non-classical states in the presence of many
uncontrollable environmental degrees of freedom. Advancements in controlling
quantum devices highlight the role of dissipation engineering in quantum
error correction. The mechanism of dissipation is embedded into some
systems and thus impossible to avoid entirely, even when the external
environment does not exist or can be sufficiently suppressed.

Dissipative behavior is usually generated by coupling the original
system with external degrees of freedom. Driven-dissipative many body
systems have been realized experimentally by coupling trapped ultra-cold
atoms to the optical modes of a laser-driven dispersive cavity \citep{ritsch2013,rai2015,klinder2015,zupancic2019,ferri2021,li2021}.
An increase of interest in theoretical descriptions of such systems
has followed. Counterintuitively, within the right parameter range,
dissipation can enhance coherence and entanglement \citep{plenio1999,plenio2002,beige2000,Daley2009,joshi2013,ates2012}.
This stabilization leads to a wealth of interesting phenomena, including
emergent phase transitions, many body pair coherent states, and novel
mode competition and symmetry breaking. In two-photon driven bosonic
lattice models, the dissipative steady states can be found exactly
\citep{roberts2023}. A two-particle loss term can increase correlations
to the point of effectively inhibiting dissipation altogether \citep{kiffner2011}.
In high Tc superconductors, non-local dissipative bosonic mediators
can act coherently and increase the superconducting critical temperature
$T_{c}$ \citep{setty2019}. The stabilizing effect of dissipation
can also facilitate experimental observation of non-equilibrium and
exotic states, such as superfluid time crystals \citep{kessler2020,kessler2021,chen2021,scarlatella2019}.
Bosonic pairs, or doublons, have been studied in systems with loss,
including three-body losses, which can be used to realize effective
three-body interactions \citep{mark2020}. The complex nature of driven-dissipative
many body models means that it is not possible to fully describe them
using methods that do not account for quantum fluctuations and information
on the spatial distributions of individuals \citep{shchesnovich2010,roberts2023}.
Therefore, up to now, the body of work has consisted mainly of relatively
limited approaches, such as few-body systems and one-dimensional studies
\citep{bonnes2011}. Furthermore, for all the new and interesting
phenomena that have already been observed, dissipation has consistently
been treated as an external factor. In this work, we focus on a different
facet of dissipative behavior: one that is an implicit property of
a strongly correlated model with extended interactions. It is known
that dissipation can generate effective many body interactions; we
show that the opposite is also possible: many body interactions can
themselves be a source of dissipative behavior.

The generic Bose-Hubbard model (BHM) for strongly interacting bosons
has been studied using a plethora of methods and approaches \citep{Gersch1963,CapogrossoSansone2007,CapogrossoSansone2008,Greiner2002,JimenezGarcia2010,Mazzarella2006,Ohliger2013,Sajna2015}.
Its extended versions are much more laborious to analyze and therefore
less abundant. It is not enough to study simplifications of extended
BHMs, as those cannot describe quantum fluctuations, especially in
lower dimensions \citep{Stasinska2021,Travin2017,Jiang2012,Hatsugai1990,Eckholt2009,Berciu2010,PhysRevLett.99.206401}.
Failure to capture the long length scale averages of order parameters
near critical points leads to unphysical phase transitions for arbitrary
chosen densities. The complications extend to experiments: pure density-induced
tunneling is difficult to explicitly replicate in bulk materials,
due to its complexity and lack of control over experimental parameters.
Nevertheless, many body correlations are always present in optical
lattice systems, even if only the standard Hamiltonian is used to
analyze experimental data \citep{Dobrzyniecki2016,Dobrzyniecki2018a}.
Whatever methods are used should not exclude correlations from the
start. 

The main motivation of the presented work are both previous theoretical
considerations and experimental data. The behavior of the condensate
is complex at low temperatures, because of intricate interactions
that come into play in such conditions. A depletion (the existence
of a finite non-condensed fraction) arises from quantum fluctuations
and affects the coherence. Quantum fluctuations generate interactions
which are not present explicitly in the model, such as the density
induced interaction. The latter can be derived separately, as an extended
version of the generic Bose-Hubbard Hamiltonian. Such interactions
are usually understood as factors which merely support the coherence
of the condensed fraction of atoms \citep{Eckholt2009}. However,
that is only true when the approach excludes quantum fluctuations.

Path integrals used in this paper constitute a flexible framework
beyond the limitations of those simpler approaches that fail to reproduce
quantum fluctuations of collective motion. The quantum rotor method
shifts the correlations between particles from bosonic fields $b$
to phase fields $\phi$, facilitating analytical study of the critical
behavior of strongly-correlated many body systems. These methods allow
to observe that pair condensation occurs implicitly within the density-induced
tunneling (DIT) BHM \citep{KRZYWICKA2022168973}. It is represented
by a double cosine term in the effective phase action. Double cosine
terms can be further linked to dissipative behavior \citep{PhysRevB.72.014509}.
The pairing term that emerges from density-induced tunneling can therefore
be treated as dissipative. We study the effect such dissipation has
on the single-particle superfluid.

We consider two versions of a pairing-based dissipative model derived
from the DIT BHM. One is assumed as a coupling between the standard
single condensate and the pair condensate, which is treated as an
external source \citep{Simanek1994a}. The other version is derived
directly from the imaginary time--dependent effective phase model
and contains an intrinsic dissipative term. We compare the critical
lines of the two systems and study the effect of the pair condensate
on the single condensate at different particle densities. We observe
a revival of the single superfluid as the DIT coefficient, which generates
the pairing mechanism and therefore the dissipative behavior, increases.

We would like to emphasize the difference between our approach and
previous ones, including several types of dissipation--like processes
that can be found in the literature. The dissipative, or open, quantum
system consists of two elements: a main, closed system is coupled
to an often (but not necessarily) larger classical or quantum system.
The environment can be assumed to have Markovian-like properties;
its dynamics can be described using the master equation in its Lindblad
form \citep{PhysRevA.88.043635,PhysRevA.97.013853,poletti,kordas}.
The latter strongly depends on the nature of the coupling between
systems. This approach is very prolific, since it can be applied to
various experimentally relevant processes, depending on the form of
the Lindblad operator. Importantly, although the Lindblad operator
conserves the total particle number, it destroys coherence throughout
the entire system. A quantitative determination of the effects of
dissipation in many body systems is possible only in terms of a fully
quantum mechanical description of the model \citep{Leggett2001,Simanek1994a}.
All the environmental modes which give rise to relevant dissipation
mechanisms have to be included in such considerations. A common practice
is to assume the nature of the coupling; environmental modes are usually
represented as harmonic oscillators with a continuously distributed
resonant frequency. One version of our calculations makes use of this
approach; we show its benefits and drawbacks. In the literature, the
coupling which provides dissipative effects is always assumed in the
microscopical model and any related macroscopic phenomena emerge from
that assumption. We derive dissipative behavior from within the microscopic
model itself. We explore how the condensate might be affected by an
intrinsic type of dissipation, as opposed to one inserted into the
Hamiltonian via a separate, external term.

The closest experimental setup to realize this model can be constructed
within optical lattices and has in fact already been realized \citep{Tenart2021},
including liquid Helium experiments \citep{article,PhysRevLett.117.235303}.
Considerable interest since the realization of atomic condensates,
especially in the context of quantum depletion, has led to counterintuitive
observations in systems of differing interaction strength and gas
dilution \citep{PhysRevLett.125.165301,PhysRevLett.96.180405}. In
systems with strong interactions, the depletion is large in comparison
with weakly-interacting diluted quantum gases. In experiments with
liquid Helium, the fraction of correlated particle pairs coexists
with the ground state of macroscopically occupied condensate. Furthermore,
in the high density and strongly-interacting regime, pairs with anti-correlated
momenta were detected \citep{Tenart2021}. 

We derive a system of relatively low densities and strong interactions.
Experimental realization would require control not only of the extended
interaction, but also of the amount of energy required to add or subtract
particles from the system: and therefore the density of the particles.
We show that suppressing density fluctuations is the key starting
point to implementing this extended Bose-Hubbard model in experiments.
Our calculations can help understand how the condensate is affected
by correlations that emerge from the system itself in experiments
with strongly-interacting Bose atoms. We thus provide a natural explanation
of coherence loss within quantum systems, which are not necessarily
connected with external degrees of freedom, but themselves generate
such dissipative environments.

\section{Model}

The theoretical description of strongly interacting bosons placed
in a two-dimensional square lattice starts with the Bose-Hubbard model
(BHM) with density-induced tunneling (DIT):
\begin{align}
\hat{H}= & \frac{U}{2}\sum_{i}\hat{n}_{i}\left(\hat{n}_{i}-1\right)-\frac{1}{2}t\sum_{\left\langle i,j\right\rangle }\hat{a}_{i}^{\dagger}\hat{a}_{j}-\mu\sum_{i}\hat{n}_{i}+\nonumber \\
 & -T\sum_{\left\langle i,j\right\rangle }\hat{a}_{i}^{\dagger}\left(\hat{n}_{i}+\hat{n}_{j}\right)\hat{a}_{j}+c.c.,\label{eq:dit-hamiltonian}
\end{align}
with on-site interaction $U$, nearest-neighbor tunneling $t$, chemical
potential $\mu$ and density-induced tunneling $T$. We concentrate
on the low temperature limit, and low densities. These assumptions
surface naturally during analysis and will be explained further as
they become relevant. In low temperatures, three phases can be recognized
in this system: the Mott insulator phase, in which particles occupy
lattice sites evenly and coherence is lost, the single particle superfluid
and the pair superfluid. We focus specifically on the impact of the
pair condensed fraction on coherence in the single particle condensed
phase.

\section{Method}

The quantum rotor (QR) analysis, used to prevent the $\mathrm{U}\left(1\right)$
symmetry of the variables, is divided into two parts. First, two sets
of coefficients are determined for the effective phase model. Further
treatment is the same for both options: the obtained phase model is
mapped onto the quantum rotor model. This method reduces the problem
of calculating critical lines to finding the saddle point of the rotor
constraint. To concentrate on the changes resulting from analyzing
a new physical system, all unnecessary details of the calculations
are omitted. It is worth emphasizing that although the approach is
known, its application to a new Hamiltonian is rather challenging,
since the model is complex and its critical properties are governed
by the preserved phase correlations.

\subsection{$\mathrm{U}\left(1\right)$ description of the model}

Using the QR method within the path integral framework, the DIT BHM
can be rewritten as a phase model \citep{KRZYWICKA2022168973}. This
requires gauge transformation, which introduces the phase field $\phi$
and changes the bosonic variables:
\begin{align}
a_{i}\left(\tau\right) & =e^{i\phi_{i}\left(\tau\right)}b_{i}\left(\tau\right),\\
\bar{a}_{i}\left(\tau\right) & =e^{-i\phi_{i}\left(\tau\right)}\bar{b}_{i}\left(\tau\right).
\end{align}
A $4\times4$ Nambu-like space is also introduced, in order to express
the amplitudes $b$ in terms of a Gaussian integral. After integration,
the partition function of the obtained phase-only model is:
\begin{align}
\mathcal{Z}= & \int\mathcal{D}\phi\,e^{-\mathcal{S}\left[\phi\right]},
\end{align}
with effective action
\begin{align}
\mathcal{S}\left[\phi\right] & =\int_{0}^{\beta}d\tau\,\left\{ \sum_{i}\frac{1}{2U}\left[\dot{\phi}_{i}\left(\tau\right)\right]^{2}+\sum_{i}\frac{\tilde{\mu}}{iU}\dot{\phi}_{i}\left(\tau\right)-\textrm{Tr}\ln\Gamma\right\} ,
\end{align}
where the $\Gamma$ matrix has the form:
\begin{equation}
\Gamma=\left(\begin{array}{cccc}
0 & \frac{1}{2}\delta_{ij}\Delta_{i} & \frac{1}{2}\left(G_{0}^{-1}+S_{ij}\right) & 0\\
\frac{1}{2}\delta_{ij}\bar{\Delta}_{i} & 0 & 0 & 0\\
0 & 0 & 0 & \frac{1}{2}\delta_{ij}\Delta_{i}\\
0 & \frac{1}{2}\left(G_{0}^{-1}+S_{ij}\right) & \frac{1}{2}\delta_{ij}\bar{\Delta}_{i} & 0
\end{array}\right)
\end{equation}
and parameters present within are:
\begin{equation}
G_{0}^{-1}=\left(\frac{\partial}{\partial\tau}+\bar{\mu}\right),\label{eq:G0-og}
\end{equation}
\begin{align}
S_{ij}= & -\left(t-2T\right)e^{-i\phi_{ij}\left(\tau\right)}-\frac{4\bar{\mu}}{U}Te^{-i\phi_{ij}\left(\tau\right)}+\nonumber \\
 & -\frac{8}{U}T^{2}e^{-i2\phi_{ij}\left(\tau\right)}\cdot\left(4\left\langle \bar{b}_{i}b_{j}\right\rangle +\delta_{ij}\right),\label{eq:bibj-in}
\end{align}
\begin{align}
\Delta_{i} & =-\frac{8}{U}T^{2}e^{-i2\phi_{ij}\left(\tau\right)}\left\langle b_{i}b_{i}\right\rangle ,\\
\bar{\Delta}_{i} & =-\frac{8}{U}T^{2}e^{-i2\phi_{ij}\left(\tau\right)}\left\langle \bar{b}_{i}\bar{b}_{i}\right\rangle ,\label{eq:bibj-fin}
\end{align}
where the chemical potential has been shifted to
\begin{align}
\bar{\mu} & =\mu+\frac{U}{2}.
\end{align}

This phase model corresponds exactly to the DIT BHM, with no approximations
required. Following standard procedure, the next step is simplifying
the effective action into a manageable form. The trace of the Green's
function can be rewritten and approximated by $\ln\left(1+x\right)\approx x$,
resulting in
\begin{align}
\mathrm{Tr}\ln\Gamma^{-1} & \approx G_{0}^{2}\left[\bar{\Delta}_{i}\Delta_{i}-S_{ij}^{2}\right]+2S_{ij}G_{0}.
\end{align}
Calculating the bosonic averages in Eqs. (\ref{eq:bibj-in}-\ref{eq:bibj-fin})
and transforming $G_{0}$ into a more useful form completes the analytical
transformation. The effective phase action in its final form,

\begin{equation}
\mathcal{S}\left[\phi\right]=\mathcal{S}_{\mathrm{U}}\left[\phi\right]+\mathcal{S}_{1}\left[\phi\right]+\mathcal{S}_{\mathrm{2}}\left[\phi\right],\label{eq:action}
\end{equation}
is comprised of three parts: an interaction part,

\begin{equation}
\mathcal{S}_{\mathrm{U}}\left[\phi\right]=\frac{1}{2U}\sum_{\left\langle i,j\right\rangle }\int_{0}^{\beta}d\tau\left(\frac{\partial\phi_{i}}{\partial\tau}\right)^{2}+\frac{\bar{\mu}}{iU}\dot{\phi}_{i}\left(\tau\right),
\end{equation}
a single condensation part,
\begin{equation}
\mathcal{S}_{1}\left[\phi\right]=g_{1}\sum_{\left\langle i,j\right\rangle }\int_{0}^{\beta}d\tau\cos\left[\phi_{i}\left(\tau\right)-\phi_{j}\left(\tau\right)\right],\label{eq:cos1term}
\end{equation}
and a pair condensation part,

\begin{equation}
\mathcal{S}_{\mathrm{2}}\left[\phi\right]=g_{2}\sum_{\left\langle i,j\right\rangle }\int_{0}^{\beta}d\tau d\tau^{'}\,\cos2\left[\phi_{i}\left(\tau\right)-\phi_{j}\left(\tau^{'}\right)\right].\label{eq:cos2term}
\end{equation}
The condensate coefficients $g_{1}$ and $g_{2}$ depend on the treatment
of $G_{0}$. We consider two possible approaches. The simpler option
is to approximate $G_{0}$. Deriving the coefficients explicitly using
Eq. (\ref{eq:G0-og}) is more complicated, but we find that doing
so reveals implicit dissipative behavior contained within the model.

\subsubsection{Effective amplitudes}

The traditional approach is to approximate $G_{0}$ by $b_{0}^{2}$,
which is obtained by minimizing the Hamiltonian \citep{Polak2007,Polak2009b}:

\begin{equation}
\frac{\partial}{\partial b_{0}}\left.\mathcal{H}\right|_{b=b_{0}}=0.
\end{equation}
In the case of the DIT BHM Hamiltonian, Eq. (\ref{eq:dit-hamiltonian}),
\begin{equation}
b_{0}^{2}=\frac{z\left(t-4T\right)+\left(\frac{U}{2}+\mu\right)}{U-8zT}.\label{eq:b02}
\end{equation}
This approximation has been deemed sufficient to study low-temperature
effects. For correlations in the $\boldsymbol{k}$ space, it can be
extended using e.g. the Bogoliubov approach \citep{Zaleski2011}.
Atom-atom correlations and time of flight images can thus be obtained
in optical lattice systems within one consistent theory.

The single and pair condensation coefficients, respectively, are as
follows:
\begin{align}
g_{1}= & -\frac{z\left(t-4T\right)+\left(\frac{U}{2}+\mu\right)}{U-8zT}\left(2\left(t-2T\right)+\frac{8\bar{\mu}}{U}T\right)+\nonumber \\
 & +\left[\frac{z\left(t-4T\right)+\left(\frac{U}{2}+\mu\right)}{U-8zT}\right]^{2}\left(\frac{64\bar{\mu}}{U^{2}}T^{3}+\frac{16}{U}JT^{2}\right)\times\nonumber \\
 & \quad\left\{ 2\left[\coth\left(-\frac{\beta\mu}{2}\right)+\coth\left(\frac{\beta\left(\mu+U\right)}{2}\right)\right]+1\right\} ,\label{eq:g1-1}
\end{align}
\begin{align}
g_{2}= & \left(\frac{z\left(t-4T\right)+\left(\frac{U}{2}+\mu\right)}{U-8zT}\right)^{2}\times\nonumber \\
 & \left[\left(t-2T\right)^{2}+\left(\frac{4\bar{\mu}}{U}T\right)^{2}+2\left(t-2T\right)\frac{8\bar{\mu}}{U}T\right].\label{eq:g2-1}
\end{align}
These amplitudes are known to provide adequate results in the study
of low temperature properties, e.g., to analyze the thermodynamical
functions and recover the well known $\lambda$ peaks in the specific
heat, which signal single and pair condensation phases of matter \citep{KRZYWICKA2022168973}.

\subsubsection{Derived model}

We introduce an alternative, more robust approach: keeping the original,
imaginary time--dependent form of $G_{0}$, Eq. (\ref{eq:G0-og})
which after Fourier transform takes the form of
\begin{equation}
G_{0}=\frac{-i\omega_{m}+\bar{\mu}}{\omega_{m}^{2}+\bar{\mu}^{2}}.\label{eq:G0}
\end{equation}
The condensate coefficients depend on imaginary time, providing new
physical effects:
\begin{align}
g'_{1}\left(\omega_{m}\right)= & -\frac{-i\omega_{m}+\bar{\mu}}{\omega_{m}^{2}+\bar{\mu}^{2}}\left(2\left(t-2T\right)+\frac{8\bar{\mu}}{U}T\right)+\label{eq:g1-sinh}\\
 & +\frac{1}{\left(-i\omega_{m}+\bar{\mu}\right)^{2}}\left(\frac{64\bar{\mu}}{U^{2}}T^{3}+\frac{16}{U}JT^{2}\right)\times\label{eq:g1-sinh-2}\\
 & \quad\left\{ 2\left[\coth\left(-\frac{\beta\mu}{2}\right)+\coth\left(\frac{\beta\left(\mu+U\right)}{2}\right)\right]+1\right\} ,\label{eq:g1-dissipative}\\
g'_{2}\left(\omega_{m}\right)= & \frac{1}{\left(-i\omega_{m}+\bar{\mu}\right)^{2}}\times\nonumber \\
 & \left[\left(t-2T\right)^{2}+\left(\frac{4\bar{\mu}}{U}T\right)^{2}+2\left(t-2T\right)\frac{8\bar{\mu}}{U}T\right].
\end{align}
In this version, imaginary time--dependent terms are present in both
condensation parts of the effective phase model, $\mathcal{S}_{1}$
and $\mathcal{S}_{\mathrm{2}}$. The single coefficient $g'_{1}$
generates two contributions, one of which has an additional dissipation-like
impact, Eq. (\ref{eq:g1-sinh}). However, this term depends on higher
orders of $T/U$ than $g'_{2}$, so at $T/U\ll1$ the pair dissipation
is much stronger. The second Eq. (\ref{eq:g1-sinh-2}), is negligible
in low temperatures after Matsubara summation. Therefore, in this
work, we forgo the marginally relevant contributions introduced by
the single condensation coefficient $g'_{1}$ and replace it with
the approximated $g_{1}$ of Eq. (\ref{eq:g1-1}), focusing on the
properties of the pair term, Eq. (\ref{eq:cos2term}), in low temperatures.

The effective action is much the same as in the simpler model, Eq.
(\ref{eq:action}), the only difference being that $\mathcal{S}_{\mathrm{2}}$
is now explicitly dissipative:

\begin{equation}
\mathcal{S}'_{\mathrm{2}}\left[\phi\right]=g'_{2}\sum_{\left\langle i,j\right\rangle }\int_{0}^{\beta}d\tau d\tau^{'}\,\frac{1}{\left(\tau-\tau^{'}\right)^{2}}\cos2\left[\phi_{i}\left(\tau\right)-\phi_{j}\left(\tau^{'}\right)\right],\label{eq:cos2term-1-1}
\end{equation}
where
\begin{equation}
g'_{2}=\left(t-2T\right)^{2}+\left(\frac{4\bar{\mu}}{U}T\right)^{2}+2\left(t-2T\right)\frac{8\bar{\mu}}{U}T\label{eq:g2-2}
\end{equation}
is the derived pair condensate coefficient.

\subsection{Dissipative phase models}

In many body effective phase models, dissipative terms are proportional
to $\left(\tau-\tau^{'}\right)^{-2}$. Traditionally, those terms
are added to the Hamiltonian as arbitrary external factors. In this
model, however, the microscopic Hamiltonian already contains the relevant
term. Both versions of the pair condensation part of the effective
action can be rewritten as dissipative. Pair condensates have been
shown to exhibit dissipative behavior in experiments \citep{Tenart2021,article,PhysRevLett.96.180405},
causing single condensate depletion.

Since the action derived from Matsubara time contains full information
about quantum fluctuations, the dissipative nature of the pair condensate
emerges naturally in Eq. (\ref{eq:cos2term-1-1}). After series expanding
the double cosine, we rewrite the derived pair effective action term
$\mathcal{S}'_{\mathrm{2}}$ as an explicitly dissipative term:
\begin{equation}
\mathcal{S}'_{\mathrm{2}}\left[\phi\right]=2g'_{2}\sum_{\left\langle i,j\right\rangle }\int_{0}^{\beta}d\tau d\tau^{'}\,\frac{1}{\left(\tau-\tau^{'}\right)^{2}}\left[\phi_{i}\left(\tau\right)-\phi_{j}\left(\tau^{'}\right)\right]^{2}.\label{eq:cos2term-1}
\end{equation}
In the simpler version of the model, based on Eq. (\ref{eq:b02}),
the imaginary time factor does not emerge naturally. To study the
dissipative effect of the pair term in Eq. (\ref{eq:cos2term}), we
treat the two condensates as separate, harmonically coupled systems:
condensed bosons submerged in a bath of harmonic potential, created
by the pair condensed system. The derivation of the effective action
is typical for such many body systems and has been carried out under
various circumstances \citep{Leggett2001}. The double cosine action
term in Eq. (\ref{eq:cos2term}) is then transformed into a dissipative
term:
\begin{equation}
\mathcal{S}_{\mathrm{2}}\left[\phi\right]=2g_{2}\sum_{\left\langle i,j\right\rangle }\int_{0}^{\beta}d\tau d\tau^{'}\,\left[\frac{\phi_{i}\left(\tau\right)-\phi_{j}\left(\tau^{'}\right)}{\tau-\tau^{'}}\right]^{2}.\label{eq:cos2term-1-2}
\end{equation}

Ultimately, the two approaches differ only by their pair condensate
coefficients: 

\begin{equation}
G_{0}\begin{array}{c}
\nearrow\\
\searrow
\end{array}\begin{array}{c}
b_{0}\begin{array}{cc}
\text{coupled condensates} & \begin{array}{l}
\rightarrow g_{1}\text{ single particle}\\
\rightarrow g_{2}\text{ pair }
\end{array}\end{array}\\
\\
G_{0}\begin{array}{cc}
\text{full treatment} & \begin{array}{cl}
\rightarrow g'_{1} & \rightarrow g_{1}\text{ single particle}\\
 & \rightarrow g'_{2}\text{ pair }
\end{array}\end{array}
\end{array}
\end{equation}
At a glance, the difference is trivial, but the two models exhibit
substantially distinct behavior, as shown in the Results section.
The proper treatment of quantum fluctuations requires an understanding
of the properties of the derived actions, as well as the application
of relevant approximations, which are different for the assumed and
the derived model.

\subsection{Quantum rotor model mapping}

The two models presented in the previous section describe the same
phenomenon, but emerge from different interactions in two different
systems, and as such are ruled by different pair coefficients $g_{2}$.
However, the distinction only becomes relevant after the critical
line equation has been derived. Thus, in the following section, both
versions of the dissipative phase model can be treated identically,
as one. The effective phase model is mapped onto the quantum rotor
model. The free energy of the latter is then minimized with use of
the saddle point method, in order to obtain the critical line equation.
Only at this point do the two versions require separate treatment.

The Fourier--transformed quantum rotor partition function is

\begin{equation}
\mathcal{Z}=\int_{-i\infty}^{+i\infty}\left[\prod_{i}\frac{\mathcal{D}\lambda\left(\tau\right)}{2\pi i}\right]e^{-N\phi\left[\lambda\right]},
\end{equation}
where

\begin{equation}
\phi\left[\lambda\right]=-\beta\lambda-\frac{1}{2N}\sum_{k}\ln\left\{ \frac{1}{\beta\pi}\left[\lambda-g_{1}\xi_{k}+\mathcal{G}^{-1}\left(\omega_{m}\right)\right]\right\} ,
\end{equation}
with Lagrange multiplier $\lambda$ and lattice constant $\xi_{k}=2\sum_{d}\cos k_{d}$. 

The critical line equation is derived by minimizing free energy with
respect to the rotor constraint $\lambda$:

\begin{equation}
\frac{\partial\mathcal{F}}{\partial\lambda}=0.
\end{equation}

After rewriting lattice dependence in terms of the density of states
function, defined as
\begin{equation}
\rho\left(E\right)=\frac{1}{N}\sum_{k}\delta\left(E-\xi_{k}\right),
\end{equation}
the critical line equation is

\begin{equation}
1=\frac{1}{2\beta}\int dE\,\sum_{m}\frac{\rho\left(E\right)}{\lambda-g_{1}E+\mathcal{G}^{-1}\left(\omega_{m}\right)},
\end{equation}
where

\begin{equation}
\mathcal{G}\left(\tau,\tau^{\prime}\right)=\exp\left[\frac{1}{\beta}\sum_{m}\frac{1-\cos\left[\omega_{m}\left(\tau-\tau^{\prime}\right)\right]}{\frac{1}{2U}\omega_{m}^{2}+4g_{2}\left\vert \omega_{m}\right\vert }\right]\label{eq:correlator}
\end{equation}
is the phase-phase correlator. In low temperatures and densities $\mu/U<\left(1-\sqrt{3}\right)2$,
the inverse of $\mathcal{G}$ can be approximated by

\begin{equation}
\mathcal{G}^{-1}\left(\tau,\tau^{\prime}\right)\approx\frac{1}{2U}\omega_{m}^{2}+4g_{2}\left\vert \omega_{m}\right\vert .
\end{equation}
At the critical point, the Lagrange multiplier $\lambda$ can be substituted
by its saddle-point value, $\lambda_{0}=g_{1}\xi_{max}$. The critical
line equation, after performing Matsubara summation in low temperatures
$\beta\rightarrow\infty$ limit, is then

\begin{align}
1= & \frac{1}{2\pi}\int d\xi\,\frac{\rho\left(\xi\right)}{g\left(\xi\right)}\left\{ \psi^{(0)}\left[\frac{\beta U}{\pi}\left(4g_{2}+g\left(\xi\right)\right)\right]\right.\label{eq:critmain}\\
 & \left.-\psi^{(0)}\left[\frac{\beta U}{\pi}\left(4g_{2}-g\left(\xi\right)\right)\right]\right\} ,\nonumber 
\end{align}
where 
\begin{equation}
g\left(\xi\right)=\sqrt{\left(4g_{2}\right)^{2}-2\frac{g_{1}}{U}\left(\xi_{max}-\xi\right)}
\end{equation}
and $\psi^{(0)}$ are digamma functions. The critical Eq. \ref{eq:critmain}
is the second pivotal point in this analysis. It contains all information
about the system, including explicitly the geometry of the bipartite
lattices, here two-dimensional square. In the low temperature limit,
digamma functions can be approximated as logarithms, leading to the
final form of the critical line equation:
\begin{align}
1 & =\frac{1}{2\pi}\int d\xi\,\frac{\rho\left(\xi\right)}{g\left(\xi\right)}\ln\left[\frac{4g_{2}+g\left(\xi\right)}{4g_{2}-g\left(\xi\right)}\right].\label{eq:criticaleqfinal}
\end{align}

Although this analysis is constrained to the low temperature limit,
it can be expanded for finite temperatures, as well as different geometries.

\section{Results}

This work focuses on low temperatures, $\beta\rightarrow\infty$,
and low density systems, $\mu/U<\left(1+\sqrt{3}\right)/2$, as the
essential phenomena take place within this parameter space. For more
information about the range of chemical potential explored in this
work, see Section \ref{sec:Appendix-A:-Chemical}. Exemplary critical
lines determined by Eq. (\ref{eq:criticaleqfinal}), which separate
the Mott insulator (MI) and superfluid (SF) phases of the single particle
condensate, are shown in Fig. \ref{fig1}. The proper energy scale
of the system must be determined. Both nearest-neighbor tunneling
$t/U$ and density-induced tunneling $T/U$ have been normalized by
the critical value $(t/U)_{\mathrm{crit}}$, which separates the MI
and SF phases in the absence of the extended interaction. The rapid
decrease of the normalized hopping $\left(t/U\right)_{N}$ is associated
with two mechanisms. The first stems from the low temperature properties
of the phase-phase correlation function \ref{sec:Appendix-A:-Chemical}.
A series expansion around the critical point shows that the density-induced
interaction $T$ both linearly suppresses the hopping amplitude and
supports particle mobility for $\left(t/U\right)_{N}$ with larger
powers of $\left(T/U\right)_{N}$:
\begin{align}
\left(t/U\right)_{N} & \simeq1-4\sqrt{2\pi}(4\mu/U+1)\left(T/U\right)_{N}\nonumber \\
 & +48\pi\left(2\mu/U+1\right)^{2}\left(T/U\right)_{N}^{2}\nonumber \\
 & -\frac{1}{2}\left(48\pi\right)^{2}\left(2\mu/U+1\right)^{4}\left(T/U\right)_{N}^{4}\nonumber \\
 & +\frac{1}{2}\left(48\pi\right)^{3}\left(2\mu/U+1\right)^{6}\left(T/U\right)_{N}^{6}+\ldots\label{eq:expansion}
\end{align}
These two effects interchange with increasing powers of the expansion
which can be missed using premature cutoff. The second decreasing
mechanism stems from the $\mathrm{U}\left(1\right)$ approach providing
complete suppression of particle mobility; this effect cannot be analytically
derived from the critical properties of the Eq. \ref{eq:criticaleqfinal}.
A sudden revival of the coherent phase is also observed. As the density
tunneling term increases, the quantum fluctuations reestablish long
range order within the system, up until rapid cutoff. At first glance,
the results from both models in Fig. \ref{fig1} seem almost identical;
the differences are clarified further on.
\begin{quotation}
\begin{figure}
\includegraphics[width=1\columnwidth]{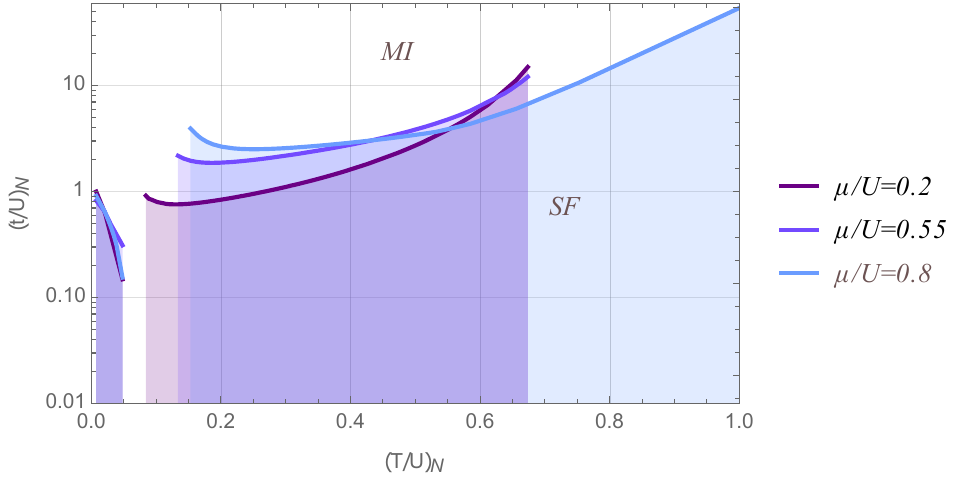}

\includegraphics[width=1\columnwidth]{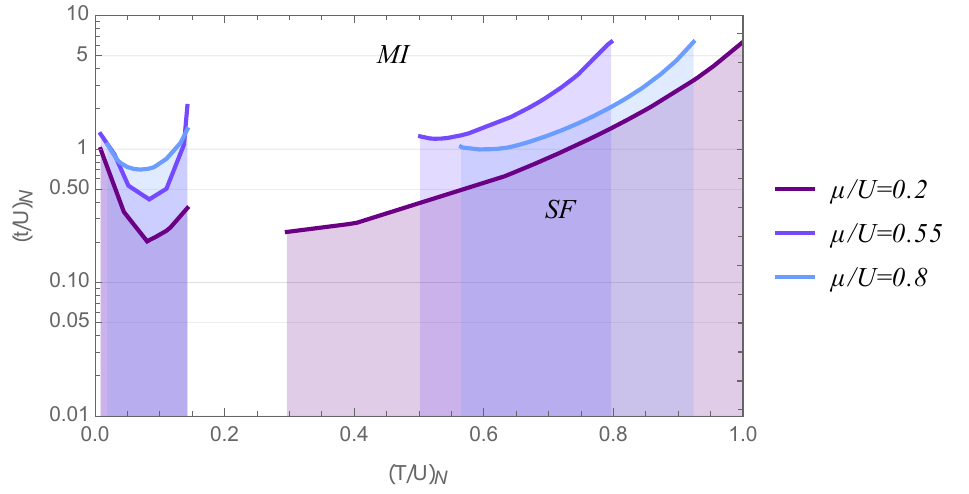}

\caption{Comparison of the dependence of normalised single hopping $\left(t/U\right)_{N}$
on the normalised (see definition in text) DIT coefficient $\left(T/U\right)_{N}$
at different chemical potentials $\mu/U$. Top: approximated model,
using $g_{2}$, Eq. (\ref{eq:g2-1}). Bottom: derived model, using
$g'_{2}$, Eq. (\ref{eq:g2-2}). The critical lines separate the superfluid
SF (below) and the Mott insulator (above) phases.}
\label{fig1}
\end{figure}
The behavior of the revival showcases the most important difference
between the assumed and derived models, presented in Fig. \ref{fig2}
in comparison to analytical results. Although the assumption made
in the simplified model about the harmonic coupling between two condensates
is reasonable and provides a qualitatively good description of the
behavior of the system, it fails to reproduce the disappearance of
coherence. It is worth noting that the quadratic potential so often
used to describe coupling between condensates cannot explain the critical
properties of the system, even though the correlation function, Eq.
\ref{eq:correlator}, has the same form in both approaches. We conclude
from Fig. \ref{fig2} that particle density is the dominant factor
in systems with the density-induced tunneling interaction. The cut-off
minimum occurs at the same value of $\mu/U$ as the tip of the superfluid--Mott
insulator lobe dominated by the density that locally conserves its
integer value. The density induced interaction could be expected to
depend strongly on the chemical potential. However, surprisingly,
the coherence restored by the density induced tunneling behaves nonmonotonically
and in opposition to the critical values of the single particle superfluid
of the generic Bose-Hubbard model. The subtlety of the phenomenon
should be also noted: the strongest coherence among the bosons is
not provided by large densities, but rather small fluctuations thereof.
The harmonic coupling model does not provide a valid description for
small densities, being almost constant throughout the relevant range
of the chemical potential values.
\end{quotation}
\begin{figure}
\includegraphics[width=1\columnwidth]{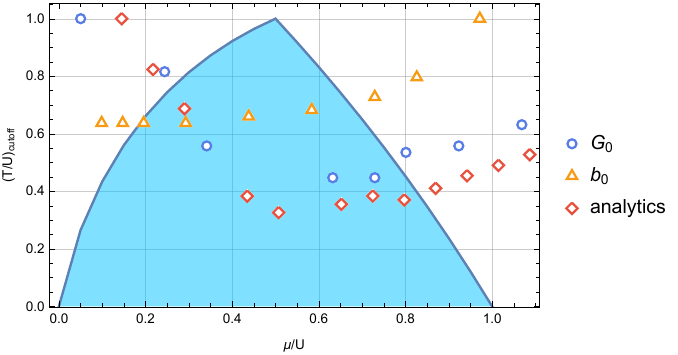}

\includegraphics[width=1\columnwidth]{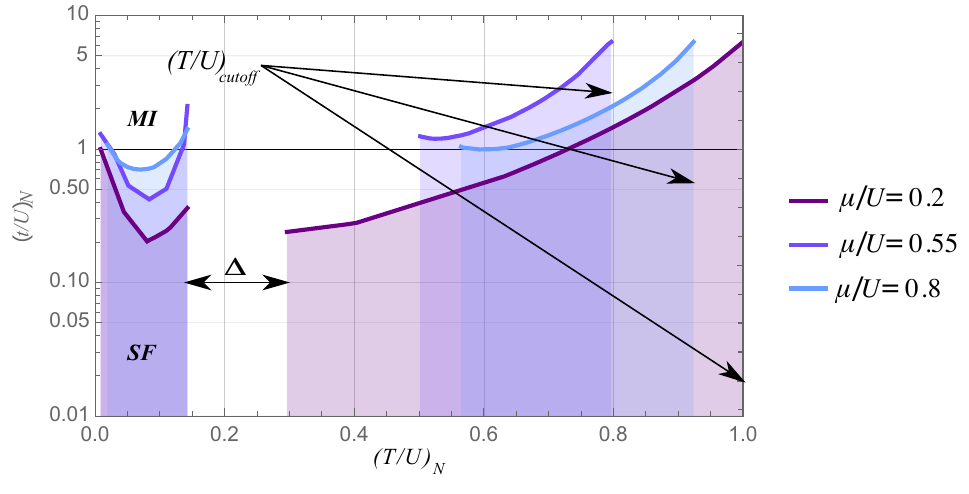}

\caption{Top: cutoff values of $(T/U)_{N}$ calculated numerically using assumed
model (orange triangles), derived model (blue circles) and analytics,
compared to the first lobe of the zero-temperature square lattice
superfluid (above) -- Mott insulator (below) phase diagram (blue
line). Bottom: visualization of the two parameters analyzed in the
top diagram and in Fig. \ref{fig3}.}

\label{fig2}
\end{figure}
It is clear that the properties of the system strongly depend on the
approach taken, but some features are shared by both of them. The
decoherence of the system and the revival of superfluidity are separated
by the gap $\Delta$, which monotonically increases with particle
density, as shown in Fig. \ref{fig3}. There are no qualitative changes
in the gap between both approaches; we conclude that it does not depend
on the character of the coupling, but rather on the quantum rotor
properties of the critical lines themselves.

\begin{figure}
\includegraphics[width=1\columnwidth]{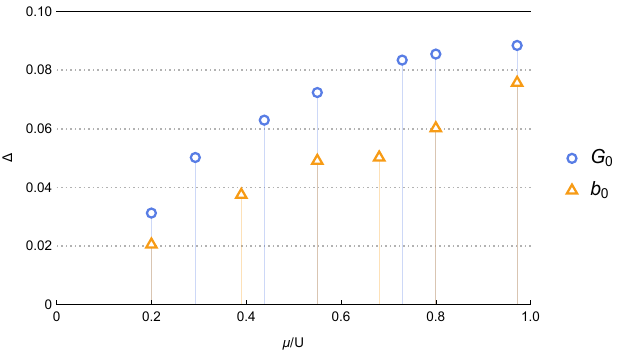}

\caption{Gap between the first decoherence breakdown and the revival of superfluidity
for both models, as indicated in the bottom diagram in Fig. \ref{fig2}.}
\label{fig3}
\end{figure}

\begin{figure}
\includegraphics[width=1\columnwidth]{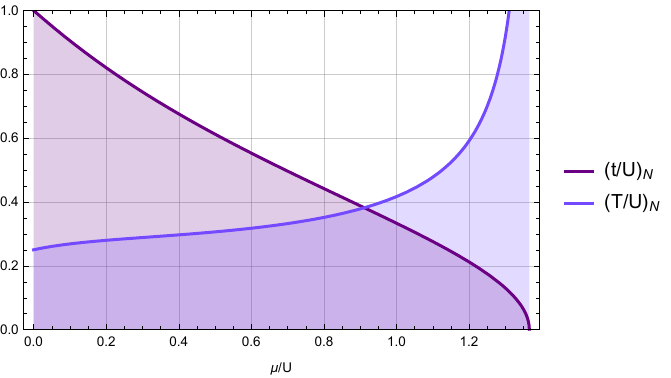}

\caption{Chemical potential critical behavior of the nearest-neighbor tunneling
and density-induced amplitudes, calculated analytically from the critical
properties of the phase-phase correlator, Eq. \ref{eq:correlatorlow}
with the upper limit $\mu/U=\left(1+\sqrt{3}\right)/2\simeq1.366$.
The diagram shows disappearance of both tunneling amplitudes for large
condensates densities. }
\label{fig4}
\end{figure}

The importance of chemical potential when density-induced tunneling
is present led us to analyze the properties of the tunneling amplitudes
relative to density. The interesting diagram in Fig. \ref{fig4} was
derived analytically from the phase-phase correlation function in
Eq. \ref{eq:correlatorlow}. Increasing particle density has different
effects on the nearest-neighbor tunneling $t/U$ and density-induced
tunneling $T/U$. The single amplitude $t/U$ counterintuitively decreases
monotonically, with a rather steep decent, and finally goes to zero.
In contrast, the DIT stays almost constant, before diverging rapidly
to infinity at high densities. The high-density critical behavior
of both amplitudes occurs at the same point of $\mu/U=\left(1+\sqrt{3}\right)/2$.
These results suggest that in systems with extended interactions,
the chemical potential governs almost all the properties of the system,
both diminishing the coherent state and at the same time supporting
correlated hopping between bosons. The magnitude of both tunneling
amplitudes is equal at $\mu/U=\left(1+\sqrt{7}\right)/4\simeq0.91$,
where $\left(t/U\right)_{N}=\left(T/U\right)_{N}=\sqrt{\sqrt{7}-5/2}\simeq0.38$.
That provides the boundary of prepotency of density induced interaction.

\section{Conclusions}

In this work, we have shown a quantum rotor analysis of the dissipative
aspects of an interaction mediated by particle density within a bosonic
system. The density-induced tunneling interaction is known to both
affect the single condensate and generate a pair condensed phase.
The two condensates can coexist and further affect each other. Our
analysis concentrated on the influence of the pair condensed fraction
on the single condensate. The system can be driven out of a dissipative
state into superfluid. A strong enough DIT interaction can boost coherence
among single particles, providing long range order. Taking quantum
fluctuations into account leads to decoherence of the system for small
values of the density induced term. This is contrary to common beliefs
that the aforementioned interaction only supports single particle
superfluidity. Observing such phenomena requires including quantum
fluctuations in theory and higher values of density induced interaction
amplitudes in experiments.

We studied the system using two different approaches. The first approach
assumed that the single superfluid is in contact with a harmonic reservoir
of pair superfluid. This version provides correct predictions of the
dissipative character of the environment; however, it fails at large
densities. We thus show that a constructed theory, which assumes harmonic
coupling between the two condensates, cannot provide a proper description
of the critical behavior of the system, even though it might to some
degree take into account quantum fluctuations. The second approach
was based on preserving the unabbreviated form of the phase-phase
correlator and its imaginary time properties. This version provides
a valid description and predicts an unprescribed cutoff of coherence
within the single particle superfluid. It is shown that particle density
governs the behavior of the system and imposes interchangeable phase
transitions. The latter might be easily missed: in theory, by assuming
harmonic character of inter--condensate coupling; in experiments,
by not controlling the density and amplitude of the density induced
interaction.

\section{Appendix A: Low temperature properties of the phase-phase correlation
function\label{sec:Appendix-A:-Chemical}}

\begin{figure}
\includegraphics[width=1\columnwidth]{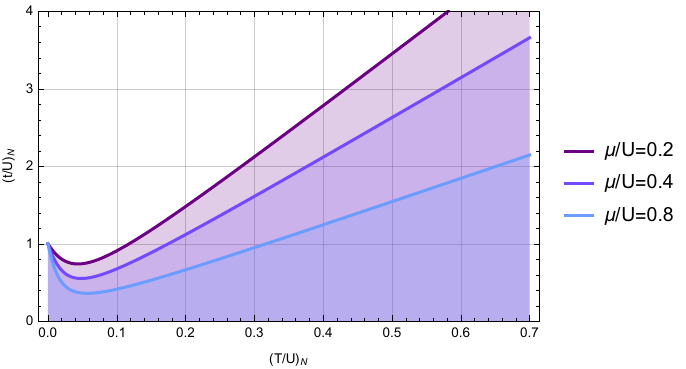}

\includegraphics[width=1\columnwidth]{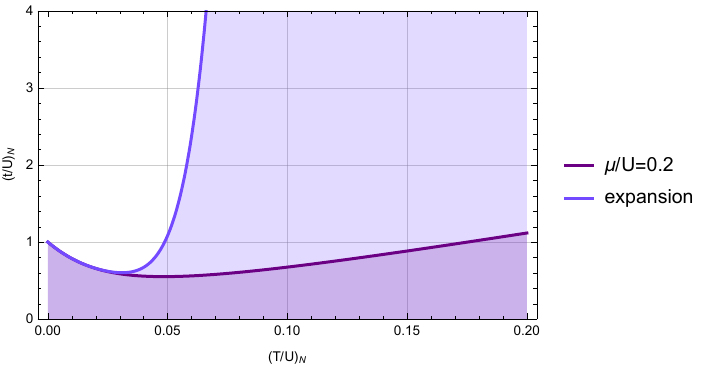}

\caption{Comparison of the dependence of normalized single hopping $\left(t/U\right)_{N}$
on the normalized DIT coefficient $\left(T/U\right)_{N}$ at different
chemical potentials $\mu/U$, resulting from analytical calculations
of the critical properties of the phase-phase correlation function
Eq. \ref{eq:correlatorlow}. Bottom: analytical model versus expansion,
Eq. \ref{eq:expansion}.}

\label{fig5}
\end{figure}

\begin{figure}
\includegraphics[width=1\columnwidth]{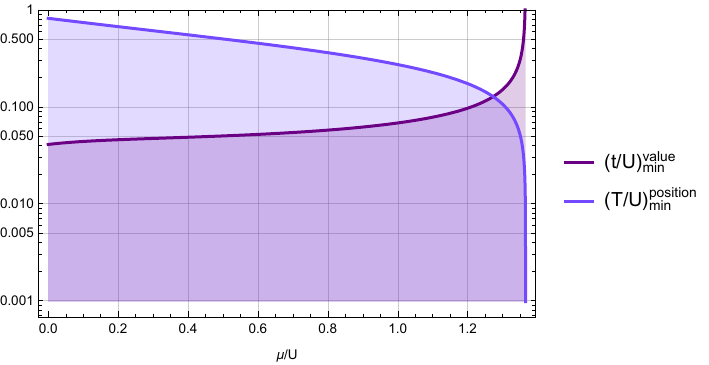}

\caption{Chemical potential dependence of the position of the minimum density
induced $\left(T/U\right)_{\mathrm{min}}$ and hopping $\left(t/U\right)_{\mathrm{min}}$
amplitudes from Fig \ref{fig5}.}
\label{fig6}
\end{figure}

After Mastubara summation and consequent Fourier transform, the full
form of the phase-phase correlator, Eq. (\ref{eq:correlator}), can
be rewritten as
\begin{align}
\mathcal{G}\left(\omega_{m}\right) & =\frac{\exp c}{\sqrt{2\pi}}b^{-\frac{a}{2}-\frac{1}{2}}\Gamma\left(\frac{a+1}{2}\right)\,_{1}F_{1}\left(\frac{a+1}{2};\frac{1}{2};-\frac{\omega_{m}^{2}}{4b}\right),\label{eq:correlatorlow}
\end{align}
where
\begin{align}
a & =\frac{2}{8\pi g_{2}},\\
b & =\frac{\pi^{2}}{24\pi\beta^{2}g_{2}},\\
c & =\frac{2H_{\frac{4g_{2}\beta}{\pi}}+2\ln\frac{2\pi}{\beta}-\frac{\pi}{\left(4\beta g_{2}+\pi\right)}}{8\pi g_{2}},
\end{align}
$\Gamma$ is the Euler gamma function, $\,_{1}F_{1}$ is the Kummer
confluent hypergeometric function, and $H_{n}$ is the $n^{\textrm{th}}$
harmonic number.

The correlator is convergent as long as $g_{2}<\left(8\pi\right)^{-1}$,
which corresponds to an upper limit on the chemical potential for
both versions of $g_{2}$, Eq. (\ref{eq:g2-1}) and Eq. (\ref{eq:g2-2}).
Within the relevant range of tunneling parameters $t$ and $T$, the
upper limit is $\mu/U=\left(1+\sqrt{3}\right)/2\simeq1.366$. We focus
on low-density systems in order to remain beneath this value. Other
properties can also be calculated from the convergence condition,
providing analytical results to compare with the numerical data obtained
from the critical line equation. Although the amplitudes $g_{0}$
and $G_{0}$ strongly affect the results, within some properties of
the system, especially where the quantum fluctuations are very strong,
the analytical predictions fit very well with numerical experiments
Fig \ref{fig5} and Fig. \ref{fig6}. The position and value of the
minimum of the normalized hopping in Fig. \ref{fig6} can be derived
analytically and yields
\begin{align}
\left(\text{\ensuremath{\frac{t}{U}}}\right)_{min}^{value} & =\frac{1}{8\sqrt{3\pi}\sqrt{-2\mu^{2}+2\mu+1}}\left|\frac{4\mu+1}{2\mu+1}\right|,\\
\left(\text{\ensuremath{\frac{T}{U}}}\right)_{min}^{position} & =\sqrt{\frac{2}{3}}\frac{\sqrt{-2\mu^{2}+2\mu+1}}{2\mu+1}.
\end{align}
The boundaries of the model parameters can be deduced from this minimum.
\begin{acknowledgments}
One of us (T.P.P.) would like to acknowledge that this work has been
done under the Maestro Grant No. DEC-2019/34/A/ST2/00081 of the Polish
National Science Centre (NCN).
\end{acknowledgments}

\bibliographystyle{apsrev4-1}
\bibliography{krzywicka}

\end{document}